\documentclass[
    aps,
    pra,
    superscriptaddress,
    showpacs,
    reprint,
    final,
    floatfix,
    twocolumn,
    showkeys,
    longbibliography,
]{revtex4-1}

\usepackage[american]{babel}
\usepackage[utf8x]{inputenc}

\usepackage{grffile} % allow for period characters in filename

\usepackage{newtxtext}
\usepackage[bigdelims,vvarbb,subscriptcorrection]{newtxmath}
\usepackage[scr=boondoxo,cal=boondoxo]{mathalfa}

\usepackage[activate={true,nocompatibility},final,tracking=true,kerning=true,spacing=true,factor=1000,stretch=20,shrink=20]{microtype}

\usepackage{bbm}% bold math
\usepackage{mathtools}
\usepackage{dsfont}% einheitsoperator
\usepackage{braket}
\usepackage{cancel} %Durchstreichen

\usepackage{graphicx}% I Fig.~files

\usepackage{siunitx} %Fuer einheiten und nicefrac
\usepackage{xfrac}

\usepackage{ragged2e}
\usepackage{array}

\usepackage{booktabs}
\usepackage{tabularx,colortbl}
\usepackage{multirow}
\usepackage[table]{xcolor}
\usepackage{adjustbox}

\usepackage{cellspace}
\newcolumntype{C}{>{\centering\arraybackslash\hspace{0pt}}X}%
\newcolumntype{L}{>{\raggedright\arraybackslash\hspace{0pt}}X}%
\newcolumntype{R}{>{\raggedleft\arraybackslash\hspace{0pt}}X}%

\newcolumntype{U}{>{\centering\arraybackslash\hspace{0pt}$}X<{$}}%
\newcolumntype{V}{>{\raggedright\arraybackslash\hspace{0pt}$}X<{$}}%
\newcolumntype{W}{>{\raggedleft\arraybackslash\hspace{0pt}$}X<{$}}%

\newcolumntype{S}{>{\columncolor[cset-aps-LightColdGray]{1}\centering\arraybackslash\hspace{0pt}$}X<{$}}%

\newcolumntype{Z}{>{$}X<{$}}

\makeatletter
\newcommand{\sbar}{{\mathchoice
  {\smash@bar\textfont\displaystyle{0.25}{1.2}S}
  {\smash@bar\textfont\textstyle{0.25}{1.2}S}
  {\smash@bar\scriptfont\scriptstyle{0.25}{1.2}S}
  {\smash@bar\scriptscriptfont\scriptscriptstyle{0.25}{1.2}S}
}}
\newcommand{\smash@bar}[4]{%
  \smash{\rlap{\raisebox{-#3\fontdimen5#10}{$\m@th#2\mkern#4mu\mathchar'26$}}}%
}
\makeatother

\definecolor{cset-aps-blueberry}{RGB}{28,128,158}
\definecolor{cset-aps-blue}{RGB}{46,44,184}
\definecolor{cset-aps-turquoise}{RGB}{0,67,88}
\definecolor{cset-aps-limegreen}{RGB}{190,219,67}
\definecolor{cset-aps-green}{RGB}{31,138,112}
\definecolor{cset-aps-yellow}{RGB}{255,225,25}
\definecolor{cset-aps-orange}{RGB}{253,116,0}
\definecolor{cset-aps-red}{RGB}{219,0,43}
\definecolor{cset-aps-violett}{RGB}{142,68,173}

\definecolor{cset-iqst-darkgray}{RGB}{99,99,99}
\definecolor{cset-iqst-mediumgray}{RGB}{165,165,165}
\definecolor{cset-iqst-lightgray}{RGB}{225,225,225}
\definecolor{cset-iqst-darkgreen}{RGB}{60,125,16}
\definecolor{cset-iqst-mediumgreen}{RGB}{112,173,71}
\definecolor{cset-iqst-lightgreen}{RGB}{169,209,142}
\definecolor{cset-iqst-darkblue}{RGB}{18,84,168}
\definecolor{cset-iqst-mediumblue}{RGB}{91,115,213}
\definecolor{cset-iqst-lightblue}{RGB}{157,195,230}
\definecolor{cset-iqst-darkorange}{RGB}{237,77,13}
\definecolor{cset-iqst-mediumorange}{RGB}{237,125,49}
\definecolor{cset-iqst-lightorange}{RGB}{244,177,131}
\definecolor{cset-iqst-darkyellow}{RGB}{255,183,0}
\definecolor{cset-iqst-mediumyellow}{RGB}{255,205,79}
\definecolor{cset-iqst-lightyellow}{RGB}{255,228,161}
\definecolor{cset-iqst-darkred}{RGB}{163,12,10}
\definecolor{cset-iqst-mediumred}{RGB}{255,66,64}
\definecolor{cset-iqst-lightred}{RGB}{255,132,130}

\definecolor{cset-material-white}{HTML}{FFFFFF}
\definecolor{cset-material-lightgray}{HTML}{BFBFBF}
\definecolor{cset-material-mediumgray}{HTML}{808080}
\definecolor{cset-material-darkgray}{HTML}{404040}
\definecolor{cset-material-black}{HTML}{000000}
\definecolor{cset-material-lightblue}{HTML}{6699FF}
\definecolor{cset-material-mediumblue}{HTML}{3366CC}
\definecolor{cset-material-darkblue}{HTML}{003399}
\definecolor{cset-material-lightgreen}{HTML}{99CC33}
\definecolor{cset-material-mediumgreen}{HTML}{00CC00}
\definecolor{cset-material-darkgreen}{HTML}{669900}
\definecolor{cset-material-yellow}{HTML}{FFCC00}
\definecolor{cset-material-lightorange}{HTML}{FF9900}
\definecolor{cset-material-darkorange}{HTML}{FF6600}
\definecolor{cset-material-red}{HTML}{CC0000}

\definecolor{cset-aps-DarkColdGray}{RGB}{61,61,70}
\definecolor{cset-aps-MediumColdGray}{RGB}{142,142,147}
\definecolor{cset-aps-LightColdGray}{RGB}{233,233,233}

\definecolor{cset-corr-user1}{RGB}{255,66,64}
\definecolor{cset-corr-user2}{RGB}{237,125,49}
\definecolor{cset-corr-user3}{RGB}{142,68,173}

\usepackage{tikz}
\usetikzlibrary{decorations}

\usepackage{hyperref}
\hypersetup{%
    colorlinks=true,
    linkcolor={cset-aps-red},
    linkbordercolor={cset-aps-red},
    filecolor={cset-aps-orange},
    filebordercolor={cset-aps-orange},
    citecolor={cset-aps-blue},
    citebordercolor={cset-aps-blue},
    urlcolor={cset-aps-green},
    urlbordercolor={cset-aps-green},
    menucolor={cset-aps-limegreen},
    menubordercolor={cset-aps-limegreen},
    breaklinks=true,
    pdfborderstyle={/S/U/W 2},
    pdfpagemode=UseOutlines,
    pdfstartpage={1},
}

\newcommand{\ii}{\mathrm{i}}
\newcommand{\ee}{\mathrm{e}}
\newcommand{\dd}{\mathrm{d}}

\begin{document}

\title[Specular mirror interferometer]{%
Proper time in atom interferometers: Diffractive vs. specular mirrors \\[1ex]
\normalsize\normalfont{Published in \href{https://journals.aps.org/pra/abstract/10.1103/PhysRevA.99.013627}{Phys. Rev. A {\bfseries 99}, 013627 (2019)}}}

\author{Enno Giese}

\email{enno.a.giese@gmail.com}
\address{Institut f{\"u}r Quantenphysik and Center for Integrated Quantum
    Science and Technology (IQ\textsuperscript{ST}), Universit{\"a}t Ulm, Albert-Einstein-Allee 11, D-89069 Ulm, Germany}

\author{Alexander Friedrich}

\address{Institut f{\"u}r Quantenphysik and Center for Integrated Quantum
    Science and Technology (IQ\textsuperscript{ST}), Universit{\"a}t Ulm, Albert-Einstein-Allee 11, D-89069 Ulm, Germany}

\author{Fabio Di Pumpo}

\address{Institut f{\"u}r Quantenphysik and Center for Integrated Quantum
    Science and Technology (IQ\textsuperscript{ST}), Universit{\"a}t Ulm, Albert-Einstein-Allee 11, D-89069 Ulm, Germany}
    
\author{Albert Roura}

\address{Institut f{\"u}r Quantenphysik and Center for Integrated Quantum
    Science and Technology (IQ\textsuperscript{ST}), Universit{\"a}t Ulm, Albert-Einstein-Allee 11, D-89069 Ulm, Germany}
    
\author{ Wolfgang P. Schleich}
\address{Institut f{\"u}r Quantenphysik and Center for Integrated Quantum
    Science and Technology (IQ\textsuperscript{ST}), Universit{\"a}t Ulm, Albert-Einstein-Allee 11, D-89069 Ulm, Germany}
\address{Hagler Institute for Advanced Study and Department of Physics and Astronomy, Institute for Quantum Science and Engineering (IQSE), Texas A{\&}M AgriLife Research, Texas A{\&}M University, College Station, TX 77843-4242, USA}

\author{Daniel~M.~Greenberger}
\address{City College of the City University of New York, New York, NY 10031, USA}

\author{Ernst~M.~Rasel}
\address{Institut f{\"u}r Quantenoptik, Leibniz Universit{\"a}t Hannover, \\
    Welfengarten 1, D-30167 Hannover, Germany}

\date{\today}

\begin{abstract}
We compare a conventional Mach-Zehnder light-pulse atom interferometer based on diffractive mirrors with one that uses specular reflection.
In contrast to diffractive mirrors that generate a symmetric configuration, specular mirrors realized, for example, by evanescent fields lead under the influence of gravity to an asymmetric geometry.
In such an arrangement the interferometer phase contains nonrelativistic signatures of proper time.
\end{abstract}

\maketitle

\section{Introduction}
The redshift controversy~\cite{Mueller10} has triggered a lively debate~\cite{Wolf10,Mueller10-reply,Wolf11,Hohensee12,Wolf12,Giulini12} about the role of proper time in atom interferometers.
Unfortunately, the discussion was focused solely on a light-pulse Mach-Zehnder interferometer (MZI) where due to the symmetry of the interferometer the proper-time difference vanishes~\cite{Greenberger12}.
However, this symmetry depends crucially on the way the mirrors change the atomic trajectory.
In the present article we propose a interferometer geometry called the specular mirror interferometer (SMI) where the proper-time difference does not vanish~\cite{Greenberger12,Lemmel14,Rauch00,Note10} due to the specific nature of the mirror.\footnotetext[10]{The crucial role of mirrors in determining the phase was already pointed out in Ref.~\cite{Greenberger12}.
Even though the claim that neutron interferometers~\cite{Rauch00} are operated with specular mirrors was challenged by Lemmel in Ref.~\cite{Lemmel14}, the conclusion that specular mirrors lead to a nonvanishing proper time is valid and underlined in the present article.}

Such configurations are crucial in studying proper time effects in atom interferometers~\cite{Zych11} experimentally~\cite{Margalit15,Zhou18,Note15}\footnotetext[15]{References~\cite{Margalit15,Zhou18} describe experiments where the effect of proper time differences between interferometer paths is simulated by different magnetic gradients.}
and might be used for other tests of the foundations of physics such as the equivalence principle~\cite{Laemmerzahl98,Rosi17}.

\subsection{The role of the mirror}
The symmetry of an MZI is intrinsically linked to the diffractive nature of the mirror pulses and has to be broken in order to observe nonvanishing proper-time contributions to the measured phase.
One possibility is the use of Ramsey-Bord\'e-type configurations~\cite{Borde08,Borde93}, where diffractive beam splitters are applied \emph{asymmetrically} to both branches.
Here, we propose an alternative geometry that relies on specular mirrors~\cite{Landragin96} inverting the incoming momentum.
When combined with the influence of gravity, the specular nature of the mirrors leads to an asymmetry that causes a proper-time contribution to the interferometer phase.

There exist several proposals to use specular reflection at evanescent fields to build a cavity for atoms and in which linear gravity is taken into account~\cite{Balykin89}.
In contrast to these ideas and Fabry-P\'erot atom interferometers~\cite{Wilkens93}, in which the atoms are localized over the length of the interferometer, we use specular mirrors not to confine atoms but to investigate the output ports of a two-branch interferometer.
For an overview of specular mirrors based on evanescent fields, we refer to Refs.~\cite{Dowling96,Grimm00}.
As an alternative to evanescent fields, strong magnetic mirrors and even permanent magnetic structures can be used for atom optics~\cite{Margalit18,Sidorov02}.
%\cite{Fortagh07} Magnetic microtraps for ultracold atoms

\subsection{Overview}
Our analysis proceeds in three steps: (i) In Sec.~\ref{sec:Semiclassical} we compare the MZI and SMI in a semi-classical description, and discuss the emergence of the total phase in the laboratory frame as well as in a frame freely-falling with the atoms, (ii) we then resort in Sec.~\ref{sec:Representation-free} to a representation-free description~\cite{Schleich13, Kleinert15} of both interferometers by introducing an operator to describe the specular reflection, and (iii) we finally study in Sec.~\ref{sec:Reflection} the reflection of a particle at an exponential potential and identify our analytical results with the specular reflection operator.
We also discuss some of the challenges of such a configuration in Sec.~\ref{sec:Challenges} before we conclude in Sec.~\ref{sec:Conclusions}.
In Appendix~\ref{sec:Appendix} we use the operator formalism introduced in Sec.~\ref{sec:Representation-free} to calculate the interference pattern of an SMI.

\section{Semi-classical description}
\label{sec:Semiclassical}
Following de Broglie~\cite{DeBroglie1924} and the path integral formulation of quantum mechanics~\cite{FeynmanHibbs1964}, a particle accumulates the phase $\Phi \equiv - mc^2\!\int\!\dd\tau /\hbar$ along its path, where $m$ is the mass of the particle, $c$ the speed of light, and $\tau$ the proper time.
In the nonrelativistic limit $\Phi$ reduces, up to a global phase factor, to the classical action which can be interpreted as a nonrelativistic signature of proper time~\cite{Borde93,Greenberger01}.
In this sense, the phase of an atom interferometer is a measure of the proper-time difference between its branches.

Since proper time is determined~\cite{Misner73} by the metric tensor, and therefore by gravity, it is mandatory to exclude electromagnetic contributions from the definition of proper time.
Consequently, the proper-time difference in an MZI vanishes~\cite{Greenberger12},
and the interferometer phase is determined solely by a laser contribution~\cite{Schleich12,Schleich13}.
As we demonstrate in the following, the SMI has a fundamentally different behavior.

\subsection{Phase contributions}
In a semi-classical description, the phase $\varphi$ of a closed atom interferometer is determined by the action integral $S\equiv\oint\! \dd t L$ along the classical trajectories of the atoms~\footnote{For potentials up to quadratic order in the position, this phase coincides with that obtained from a full quantum description of the interferometer.}, that is $\varphi\equiv S/\hbar$.
The Lagrangian $L \equiv  E_\text{kin} - V_\text{grav} - V_\text{lp}$ of a typical configuration consists of three contributions~\cite{Schleich13}:
(i) the kinetic energy $E_\text{kin} \equiv m v^2(t) /2 $, where $m$ is the mass of the atom and $v=v(t)$ its velocity component parallel to gravity;
(ii) the gravitational potential $ V_\text{grav} \equiv m g z(t)$, where $g$ is the gravitational acceleration and $z=z(t)$ is the vertical position of the atom;
and (iii) the potential applied by the laser pulses $V_\text{lp} \equiv - \sum_j (\pm 1)_j \hbar \big[k z(t)+\phi_t\big] \delta(t-t_j)$, where $(\pm 1)_j$, $t_j$, $k$ and $\phi_t$ denote the direction, time, effective wavevector of the two-photon Raman transition~\footnote{The two-photon Raman transition is driven by two counterpropagating waves so that the effective momentum transfer $\hbar k$ corresponds to the sum of their single-photon momenta.}, and phase of the $j$th pulse, respectively.

We emphasize that $V_\text{lp}$ is a branch-dependent potential, and represents an essential part of the action, as it modifies the trajectories by applying momentum kicks $\pm \hbar k$ to the atomic wave packet~\footnote{We assume that the laser is aligned in parallel with gravity so that the momentum is transferred only parallel or anti-parallel to the gravitational field. In this case, the orthogonal directions become separable and are irrelevant for the phase of the interferometer. Hence, a one-dimensional theory is sufficient, and we refer in the remainder of this article only to the position and the momentum or velocity components parallel to gravity.}.
Indeed, the decomposition $V_\text{lp}=V_\text{kick}+V_\text{phase} $ underlines that 
\begin{subequations}
\begin{equation}
    V_\text{kick}\equiv- \sum_j (\pm 1)_j \hbar k z(t)\delta(t-t_j)
\end{equation}
transfers the momentum and
\begin{equation}
    V_\text{phase}\equiv- \sum_j (\pm 1)_j \hbar\phi_t \delta(t-t_j)
\end{equation}
\end{subequations}
imprints the phase of the laser on a particular branch~\footnote{This decomposition distinguishes between the dynamics of the atomic sample caused by a time-dependent linear potential, and a phase offset which does not affect the dynamics, but nonetheless contributes to the total phase of the interferometer.}.
Each of these four energies contributes to the total phase of the interferometer $\varphi = (S_\text{kin}+S_\text{grav}+S_\text{kick}+S_\text{phase})/\hbar$.

\subsection{Mach-Zehnder interferometer}
An MZI~\cite{Kasevich91} consists of a $\pi/2$-laser pulse that acts as a beam splitter, free propagation for a time $T$ in the gravitational field, a $\pi$-laser pulse to redirect the two branches, followed by another free propagation for a time $T$, and a final $\pi/2$-laser pulse to recombine the two branches, as indicated in Table~\ref{fig:comparison} on the left.
In this arrangement, both $S_\text{kin}$ and $S_\text{grav}$ have the same magnitude $\hbar \varphi_\text{g}\equiv \hbar k g T^2$, but opposite signs~\cite{Schleich13}. Hence, the proper-time difference which is proportional to $S_\text{kin}+S_\text{grav}$ vanishes.
In this sense, the phase $\varphi_\text{MZI}$ of the MZI is determined by the laser contributions $S_\text{kick}+S_\text{phase}$.
We emphasize that these results are independent of the initial position or velocity of the atom.

With the discrete second derivative $\Delta \phi \equiv \phi_0- 2 \phi_T + \phi_{2T}$ of the laser phase, we find~\cite{Schleich13}
\begin{equation}
\label{eq:phi_MZI}
    \varphi_\text{MZI} \equiv \Delta \phi - \varphi_\text{g}.
\end{equation}
The different contributions at the various stages are listed in Table~\ref{fig:comparison} on the left~\footnote{Since the laser pulses change the trajectories, they are an essential part of the potential contributing to the action. Hence, separating them from the proper-time difference is somewhat arbitrary.}.

\begin{center}
\begin{table*}
\caption{Comparison of Mach-Zehnder (MZI, left) and specular mirror interferometer (SMI, right) in the laboratory and in the freely-falling frame (top and bottom). We depict the MZI and SMI geometries in time-space diagrams and show the individual phase contributions during the interferometer sequence in the tables below. The beam splitters at $t=0$ and $t=2T$ are conventional Raman pulses connecting ground and excited states $\ket{g}$ and $\ket{e}$ in the MZI and SMI. For the MZI, a mirror Raman pulse is applied at $t=T$. The SMI has two specular mirrors, for example realized by strongly-detuned evanescent fields, that at $t=T$ invert the momenta of each branch. The initial vertical momentum is chosen to be $\hbar k/2$ for both the MZI and the SMI. Whereas $S_\text{kin}+S_\text{grav}$ vanishes in all frames for the MZI, the SMI always has  $S_\text{kin}+S_\text{grav} = -2\hbar \varphi_\text{g}$, where we have defined $\varphi_\text{g}\equiv k g T^2$.}
\label{fig:comparison}
\includegraphics{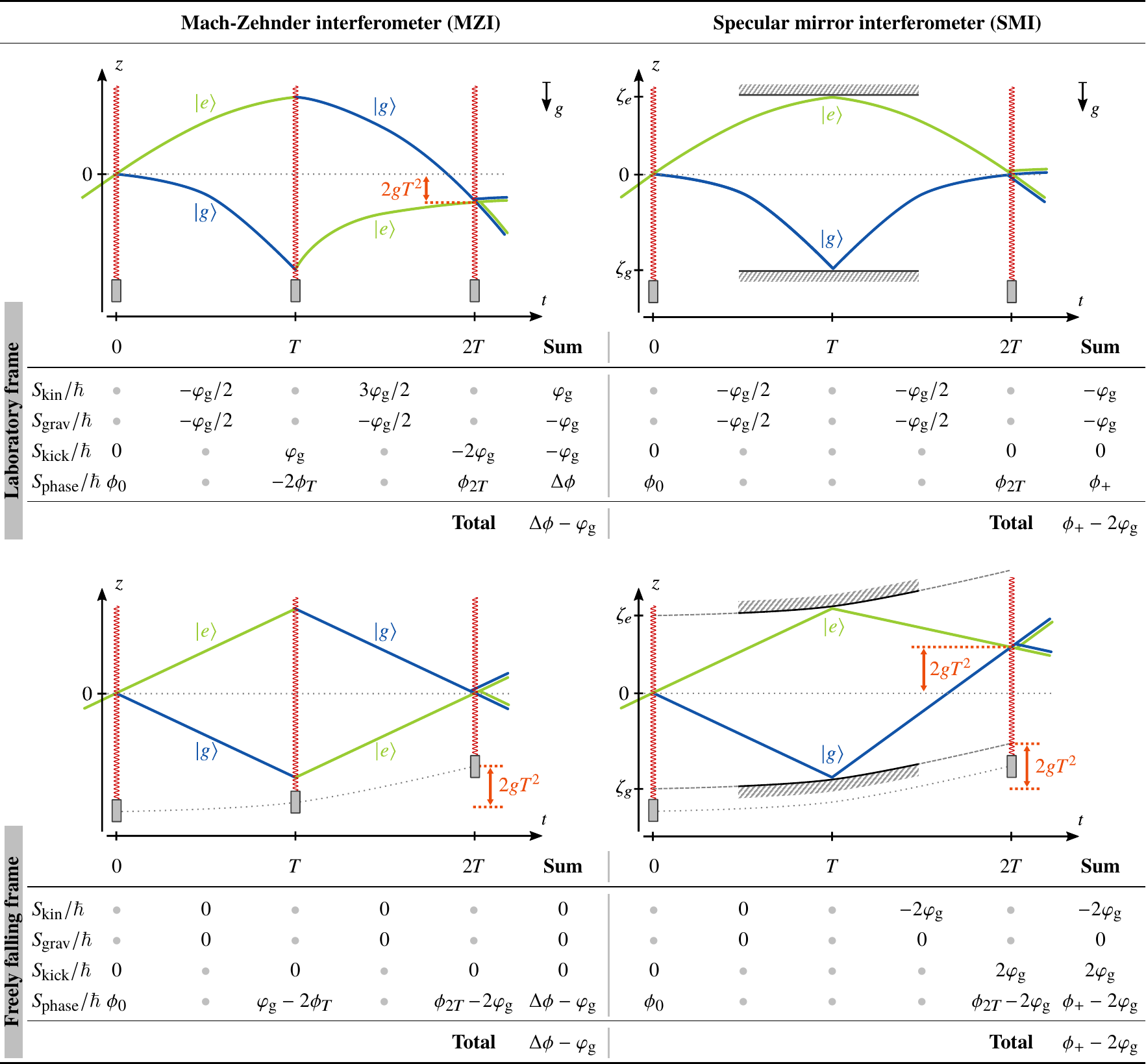}
\end{table*}

\end{center}

\subsection{Specular mirror interferometer}
Next, we consider the SMI configuration in which the $\pi$-laser pulse in the MZI sequence is replaced by a mirror that inverts the momentum of the atoms. Contrary to the diffractive mirror pulses in an MZI, which always transfer a momentum kick $\pm \hbar k$, this mirror is truly specular.

On the right side of Table~\ref{fig:comparison} we showcase the SMI where at $t=T$ the momentum is inverted.
Moreover, the positions $\zeta_e$ and $\zeta_g$ of the two mirrors are chosen in such a way that the interferometer is closed in phase space.
The figure also shows that such a geometry is symmetric with respect to the specular mirror: immediately after the final $\pi/2$-laser pulse, the atom returns to its initial height.

When we calculate the different contributions to the interferometer phase, we find in Table~\ref{fig:comparison} on the right that $S_\text{kin}+S_\text{grav}= - 2\hbar \varphi_\text{g}$ is in fact nonvanishing.
Here, we have chosen the initial vertical momentum $mv(0)\equiv \hbar k/2$ and the initial position $z(0)\equiv0$.

Hence, the total phase $\varphi_\text{SMI}$ acquired in an SMI reads
\begin{equation}
\label{eq:phi_SMI}
    \varphi_\text{SMI} \equiv \phi_+ - 2 \varphi_\text{g},
\end{equation}
where the contribution $\phi_+ \equiv \phi_0 + \phi_{2T}$ arises from the initial and final $\pi/2$-laser pulses~\cite{Giese15}.
Apart from replacing $\Delta\phi$ by $\phi_+$, $\varphi_\text{SMI}$ differs by a factor of two in $\varphi_\text{g}$ from the phase $\varphi_\text{MZI}$ of an MZI.

\subsection{Freely-falling frame}
To gain a deeper understanding of the appearance of the nonvanishing proper-time difference and the factor of two, we discuss both the MZI and the SMI in a freely-falling frame, shown in Table~\ref{fig:comparison} on the bottom.
In this particular frame the MZI is symmetric and the proper-time difference vanishes.
The laser phase is modified to $\phi_\text{f}(t)\equiv \phi_t - k g t^2/2$ in the freely falling frame; that is the laser is accelerated with respect to the atomic trajectories.
Again, only the laser leads to a phase contribution.

In contrast to the MZI, the SMI is asymmetric in the freely-falling frame because the specular mirror is accelerated: the reflection at an accelerated surface gives an additional momentum kick and the proper-time difference determined by $S_\text{kin}+S_\text{grav}$ is nonvanishing.
\clearpage
\section{Representation-free description}\label{sec:Representation-free}
\vspace{-1ex}
Next we discuss both interferometers, in a representation-free manner following Refs.~\cite{Schleich12,Schleich13,Kleinert15}.
In such a description, we do not rely on classical trajectories or the path-integral formalism, but instead use displacement operators to model the momentum transfer, the parity operator to model specular reflection, and time-evolution operators for the dynamics of the wave packet in the gravitational potential.
This treatment is independent of any representation and therefore does not imply a particular interpretation.

\subsection{Mach-Zehnder interferometer}
For simplicity, we assume that the laser drives two-photon Raman transitions in an effective two-level system between the internal states $\ket{e}$ and $\ket{g}$ of energy difference $\hbar \omega$.
We therefore describe the time evolution between the beam splitter pulses and the mirrors by the usual canonical Hamiltonian
\begin{equation}
\hat{H} \equiv \frac{\hbar \omega}{2} \big(\ket{e}\bra{e} - \ket{g}\bra{g}\big) +\frac{\hat{p}^2}{2m}+mg\hat{z},
\end{equation}
which leads us directly to the time-evolution operator ${\hat{U}(t) \equiv \exp\big(-\ii\hat{H}t/\hbar\big)}$.

The action of a Raman pulse on the atom follows from the operator
\begin{equation}
\label{eq:Raman}
    \hat{R}^{(j)}_{\pm}(t) \equiv 
     \cos{\theta_{j} \; \mathbbm{1}_\text{int}}
      -\ii\sin{\theta_{j} \, 
    \left[
        \ee^{ \pm \ii (k \hat{z} + \phi_t)}\ket{e}\!\bra{g} +
        \text{h.c.}
    \right]}
\end{equation}
where $\mathbbm{1}_\text{int} \equiv \ket{e}\!\bra{e}+\ket{g}\!\bra{g}$ and $j=B$ with $\theta_B=\pi/4$ for a beam splitter pulse, or $j=M$ with $\theta_M = \pi/2$ for a mirror pulse.

The Raman pulse does not only drive transitions between the internal states, but at the same time transfers, as a consequence of the operators $\exp{\big(\pm \ii [k\hat{z}+\phi_t]\big)}$ in~Eq.~\eqref{eq:Raman}, momentum to the atom and imprints a phase-shift on the atomic wavefunction.
To realize an SMI, we have to reverse the momentum transfer of the final Raman pulse, which we denote by the index $\pm$.

When we express the sequence of an MZI in terms of $\hat{U}$ and $\hat{R}_+$, the output postselected on the excited state is
\begin{equation}
    \ket{\psi_\text{MZI}^{(e)}} = \bra{e} \hat{R}^{(B)}_{+}(2T)\hat{U}(T)\hat{R}_{+}^{(M)}(T)\hat{U}(T)\hat{R}^{\left(B\right)}_{+}(0)\ket{\Psi_\text{in}},
\end{equation}
where the initial state $\ket{\Psi_\text{in}}$ contains both the internal and the external degrees of freedom of the atom~\footnote{In this equation and the remainder of the article we assume that the internal state is associated with one specific momentum class and therefore a semiclassical trajectory.}.

For an atom initially in the excited state $\ket{e}$ and in the state $\ket{\psi}$ describing the external degree of freedom, i.e. $\ket{\Psi_\text{in}}\equiv \ket{e}\ket{\psi}$, the probability $P_\text{MZI}^{(e)} \equiv \langle \psi_\text{MZI}^{(e)}|\psi_\text{MZI}^{(e)}\rangle $ of the atom exiting the interferometer in the excited state takes the form~\cite{Schleich13}
\begin{equation}
    P_\text{MZI}^{(e)} = \frac{1}{2}\big(1  + \cos \varphi_\text{MZI}  \big),
\end{equation}
and depends on the phase $\varphi_\text{MZI}$ given by~Eq.~\eqref{eq:phi_MZI}.

\subsection{Specular mirror interferometer}
In order to describe the SMI, we introduce the parity operator~\cite{Royer77}
\begin{equation}
\label{eq:parity}
    \hat{\Pi}(\zeta) \equiv \int \!\! \dd p \, \ee^{ \ii\, 2p \zeta / \hbar } \ket{-p}\!\bra{p}
\end{equation}
which inverts the vertical momentum at a position $\zeta$, and assume that a specular mirror acts like this operator on one particular branch.
Since we require to reflect the lower and the upper branch independently, we need two specular reflections and define the operator
\begin{equation}
\label{eq:mirror_operator}
    \hat{M}(\zeta_e,\zeta_g) \equiv- \hat{\Pi}(\zeta_e)\ket{e}\!\bra{e}-\hat{\Pi}(\zeta_g)\ket{g}\!\bra{g} .
\end{equation}
Here $\zeta_e$ and $\zeta_g$ correspond to the positions of the upper and the lower mirror, which can be associated in our geometry with the excited and ground state of the atom~\cite{Giese15}.
We specify the positions later.

For a proper comparison to the MZI, we need to project on the ground state rather than the excited state, because in the SMI we only have two instead of three pulses changing the internal state.
The momentum transfer of the final beam splitter at time $2T$ has to be reversed, as indicated by the subscript ``$-$,'' so that the atoms exciting the interferometer in the ground state are associated with an upward momentum.
We emphasize that these subtleties can be avoided using Bragg diffraction.

Hence, the state corresponding to atoms exiting the SMI in the ground state reads
\begin{equation}
    \ket{\psi_\text{SMI}^{(g)}} = \bra{g} \hat{R}^{(B)}_{-}(2T)\hat{U}(T)\hat{M}(\zeta_e,\zeta_g)\hat{U}(T)\hat{R}^{\left(B\right)}_{+}(0)\ket{\Psi_\text{in}}.
    \label{eq:state_SMI}
\end{equation}

In Appendix~\ref{sec:Appendix} we use the explicit forms of the operators corresponding to beam splitter, mirror, and time-evolution to show that the probability $P_\text{SMI}^{(g)}\equiv \langle \psi_\text{SMI}^{(g)}|\psi_\text{SMI}^{(g)}\rangle$ takes the form
\begin{equation} \label{eq:P_SMI}
    P_\text{SMI}^{(g)}=\frac{1}{2}+\frac{1}{4} \bra{\psi}\ee^{-2\ii \omega T} \ee^{\ii \tilde{\varphi}_\text{SMI}} \ee^{ 2\ii (\hat{p}-mgT)Z/ \hbar} \ket{\psi}+\text{c.c.}
\end{equation}
and depends on the initial state, where we have used the same initial condition as above.
Here, we have defined the distance $ Z \equiv \zeta_e-\zeta_g -  \hbar k T /m$, and the phase ${\tilde{\varphi}_\text{SMI} \equiv \phi_+ - \varphi_g + 2 k \zeta_g + \hbar k ^2 T /m}$ is different from~Eq.~\eqref{eq:phi_SMI}.

The phase $-2\omega T$ arises from the propagation in different internal states, and appears together with $\phi_+$ in the interference pattern like in conventional Ramsey spectroscopy.
Ideally, one would operate the interferometer at the Ramsey resonance, which might be experimentally challenging. 
However, since in our article we do not focus on this clock phase, we omit the phase $-2\omega T$ in the following but emphasize that we could have performed the complete analysis using Bragg diffraction, where the atom is always in the same internal state, and therefore, the clock phase vanishes.

When we choose $ \zeta_e-\zeta_g = \hbar k T/m$, which corresponds to the classical separation of the two mirrors, we have $Z=0$ and thus the interferometer closes in phase space~\cite{Roura14}.
In this case, the dependence on the momentum operator in Eq.~\eqref{eq:P_SMI} disappears, and the expectation value reduces to a phase factor.

Hence, we find a perfect visibility and the interference pattern for an atom initially in the excited state reduces to
\begin{equation}
    P_\text{SMI}^{(g)}= \frac{1}{2}\big(1  + \cos \tilde{\varphi}_\text{SMI} \big).
    \label{eq:P_SMI_out}
\end{equation}
When we set the position of the lower mirror to $\zeta_g = - \hbar  k T/(2m)  - gT^2/2 $, which corresponds to its classical position for $z(0)=0$ and $ v(0) = \hbar k/(2m)$, the interferometer phase $\tilde{\varphi}_\text{SMI}$ reduces to $\tilde{\varphi}_\text{SMI} = \phi_+ - 2\varphi_g \equiv \varphi_\text{SMI}$, in complete agreement with our semiclassical result from~Eq.~\eqref{eq:phi_SMI} obtained for the same initial conditions.

\subsection{Freely falling frame}
In order to compare and contrast the two interferometers in the freely-falling frame, we introduce the displacement operator
\begin{equation}
    \hat{D}(\zeta,\wp) \equiv \exp{\Big(\ii \left( \wp \hat{z}-\zeta \hat{p}\right)/\hbar \Big)}
\end{equation}
which shifts the position of a given state by $\zeta$ and its momentum by $\wp$.

Moreover, we recall the decomposition~\cite{[See{,} for example{,}~]Kajari10}
\begin{equation}
\label{e_time-evolution-expanded}
    \hat{U}(t) = \hat{D}(\zeta_t,\wp_t)\exp\Big(-\ii \hat{p}^2 t/(2m \hbar) \Big)\exp\Big(\ii m g^2 t^3/(12\hbar)\Big),
\end{equation}
corresponding to the motion in the gravitational field with $\zeta_t \equiv -gt^2/2$ and $\wp_t\equiv -mgt$. Hence, the time-evolution operator $\hat{U}$ is, up to a phase factor cubic in $t$~\cite{Zimmermann2017}, a displacement along the trajectories $\zeta_t$ and $\wp_t$ multiplied by the time evolution in the freely-falling frame.

The displacement operator $\hat{D}$ shifts the laser pulses and the mirror of the interferometer sequences to the freely-falling frame.
In fact, the transformation $ {\hat{\mathcal{R}}^{(j)}_\pm(t)=\hat{D}^\dagger(\zeta_t,\wp_t)\hat{R}_\pm^{(j)}(t)\hat{D}(\zeta_t,\wp_t)}$ of the Raman transition leads only to a replacement of $\phi_t$ in~Eq.~\eqref{eq:Raman} by ${\phi_\text{f}(t) =  \phi_t + k \zeta_t}$ as a manifestation of the acceleration of the laser with respect to the atoms.

Applying the identical transformation to the mirror operator, ${\hat{M}_\text{f}(\zeta_e,\zeta_g) = \hat{D}^\dagger(\zeta_t,\wp_t) \hat{M}(\zeta_e,\zeta_g) \hat{D}(\zeta_t,\wp_t)}$, shows that the parity operators in the definition of $\hat{M}_\text{f}(\zeta_e,\zeta_g)$ are also transformed and now take the form
\begin{equation}
    \hat{\Pi}_\text{f}(\zeta) = \int\!\!\dd p \, \ee^{2\ii(p+\wp_t) (\zeta-\zeta_t) / \hbar} \ket{-p-2\wp_t}\!\bra{p}.
\end{equation}
Here, the position $\zeta$ of the mirror is shifted by $\zeta_t$.

Moreover, an additional momentum transfer $-2\wp_t$ to the atom upon reflection arises from the acceleration of the mirror.
Due to the unitary nature of the displacement operators generating the transformation to the freely falling frame and the decomposition from Eq.~\eqref{e_time-evolution-expanded}, the output probability is identical to Eq.~\eqref{eq:P_SMI_out}.

\section{Reflection at an exponential potential}
\label{sec:Reflection}

Specular mirrors for atom optics based on a strongly-detuned evanescent electromagnetic field have already been realized experimentally in the context of the atom trampoline~\cite{Steane95,Szriftgiser96,Landragin96,Henkel99} and used to simulate the Fermi accelerator \cite{Saif98,Saif05}. Potentials enabling specular reflection of matter waves have also been implemented by means of magnetic fields \cite{Sidorov96, Hinds99, Lau99, Sidorov02} based on either current carying wires, magnetic surfaces or permanently magnetized micro-structures. Moreover, recently there have been significant advances in Stern-Gerlach interferometry with atoms \cite{Margalit18}, which has been a long standing challenge \cite{Wigner1963,Scully89,Schwinger88,Englert88,deOliveira2006,Dipankar2007}, leading to yet another method for facilitating an exponential potential. However, due to the experimental requirement for an optical access in direction of the atomic motion we focus in the following on an evanescent field mirror.

In previous implementations of evanescent-field mirrors for atomic wave packets only the lower mirror has been constructed~\cite{Abele10,Jenke11,Note11}\footnotetext[11]{It is interesting to note that recently a specular mirror was a crucial ingredient in the excitation of neutron wave packets \cite{Abele10, Jenke11} moving in the gravitational field of the earth.}.
Nevertheless, these experiments show that our assumption of an instantaneous inversion of the momentum at time $T$ is an idealized assumption and the reflection has to be realized by a steep potential with a nonvanishing interaction time, leading to additional phase contributions.

To discuss such phases, we investigate the reflection of an incoming wave from an exponentially increasing potential located at $s$.
The corresponding Schr{\"o}dinger equation reads
\begin{equation}
 -\frac{\dd^2}{\dd z^2}\psi\left(z\right)+\kappa^2\,e^{2(z-s)/\lambda}\,\psi\left(z\right)=\left(\frac{p}{\hbar}\right)^2\,\psi\left(z\right),
\label{eq:SchroedingerEq}
\end{equation}
where $\kappa\equiv\sqrt{2m V_0}/\hbar$ is the normalized strength of the mirror potential, $p$ is the incoming momentum of the wave, $m$ denotes the mass of the particle, and $V_0$ is the potential strength. We note that this simple description is only valid at positions $z<s$ which are sufficiently \cite{Henkel99,Note12} distant from the physical location $s$ of the glass plate used in the generation the evanescent field potential.\footnotetext[12]{If the atomic wave packet probes the vicinity of the glass plate during the reflection process the description in Eq.~\eqref{eq:SchroedingerEq} has to be amended. As pointed out by Henkel et al. in Ref.~\cite{Henkel99} in a first step additional potential terms due to the finite height of the potential, multiple atomic levels participating in the reflection of the atom, and surface interactions have to be taken into account.}
The parameter $\lambda$ is the decay length of the evanescent electromagnetic field and describes the steepness of the potential.

The solution of~Eq.~\eqref{eq:SchroedingerEq} is given~\cite{Henkel94,Henkel96} by Bessel functions, which reduce in the limit $z\rightarrow-\infty$ due to the vanishing potential to a superposition
\begin{equation}
    \psi(z)\propto e^{\ii p z /\hbar } - e^{2\ii \theta} e^{- \ii p z/ \hbar }
    \label{eq:WaveFunction}
\end{equation} 
of incoming and outgoing plane waves.

Thus, the asymptotic limit of the solution of~Eq.~\eqref{eq:SchroedingerEq} gives the phase
\begin{equation}
    \theta =\vartheta(p)-\big[\ln{\left(\kappa\lambda/2\right)}- s/\lambda\big]\,p\lambda/\hbar \,,
    \label{eq:WaveAfterScattering}
\end{equation}
where $\vartheta(p)\equiv \arg{\Gamma\left(1+\mathrm{i}p\lambda/\hbar\right)}$ depends on the incoming momentum $p$ and $\Gamma$ denotes the Euler-Gamma function.

Expanding the phase $\theta$ around the momentum $p_0$ allows us to write Eq.~\eqref{eq:WaveFunction} as 
\begin{equation}
    \psi(z) \propto \bra{z}\left[1 - e^{2\ii\theta_0}\hat{\Pi}(\zeta)\right] \ket{p},
\end{equation}
which is identical to the mirror operator introduced in~Eq.~\eqref{eq:parity}.

Therefore, we find in first order an effective position
${\zeta = \hbar  \vartheta^{\prime}(p_0)-\lambda\ln{\left[\kappa\lambda/2\right]}+s}$ of the mirror that depends on the position $s$ of the exponential potential and on the derivative $\vartheta^{\prime}(p_0)$, while $\theta_0\equiv \theta(p_0)-\zeta p_0/\hbar $ is a constant.
The truncation of the expansion at second order and the replacement of $\theta$ with its truncated Taylor expansion is only admissible if the second derivative of $\vartheta$ is small over the spread $\Delta p^2_0$ of an incoming wave packet, that is $\theta^{\prime\prime}(p_0)\Delta p^2_0\ll1$.

\begin{figure}
    \centering
    \includegraphics[width=\columnwidth]{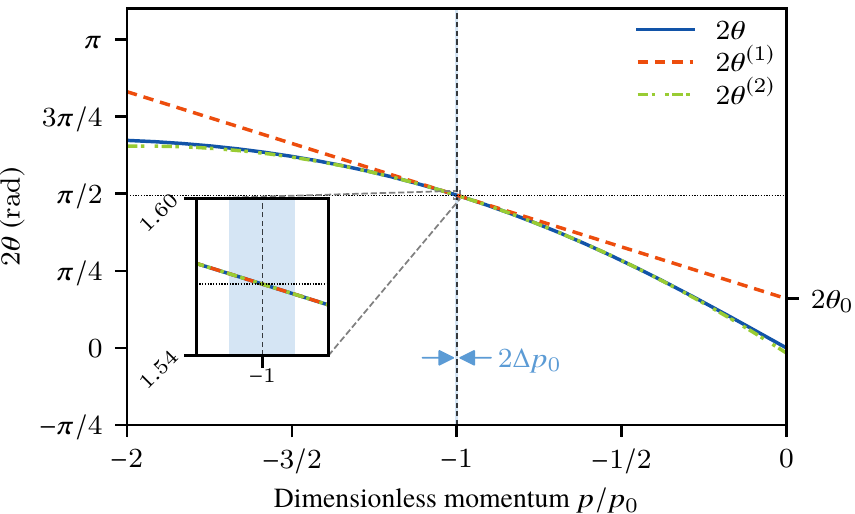}
    \caption{%
    Phase shift $\theta(p)$ imprinted onto the outgoing momentum distribution. The curves are a result of the analytic solution of the scattering problem given by ~Eq.~\eqref{eq:SchroedingerEq} for an incident wave packet. For a suitably narrow momentum distribution whose width $\Delta p_0$ fulfills $\theta''(p_0)\Delta p_0^2\ll 1$, the first-order expansion of the phase, $2(\theta_0+\zeta p/\hbar)$ (orange dashed line), is sufficient. In this regime, we can use the parity operator $\hat{\Pi}$ to describe the reflection. The second-order expansion $2\theta^{(2)}(p)$ is depicted by a green dashed-dotted line. The inset shows a detailed view of the imprinted phase shift over the width of the wave packet.
    The parameters of the exponential potential used are a potential strength of $V_0 = 20\, E_0$, where $E_0=p_0^2/(2m)$ is the initial kinetic energy of the atoms, a decay length of $\lambda = 10^{-8}\,\text{m}$ for the evanescent wave field and the location $s=0$ for simplicity. For the atoms we use ${}^{87}\text{Rb}$ and an incident momentum of ten single-photon recoil momenta $p_\text{rec}$ for the central wave packet component with respect to the $\mathrm{D}_2$-line, that is $p_0 = 10 p_\text{rec}$. The width of the atomic wave packet is set at $\Delta p_0 = 0.05 p_\text{rec}$.
    }
    \label{fig:realistic_phaseshifts}
\end{figure}
The generalization of the previous treatment to wave packets by a superposition of plane waves leads in momentum representation to $\psi(p) \propto \psi_\text{in}(p)-\psi_\text{out}(p)$.
In Fig.~\ref{fig:realistic_phaseshifts} we depict the phase shift imprinted on the outgoing momentum distribution as implied by the analytic asymptotic solution of~Eq.~\eqref{eq:SchroedingerEq}.
In particular, Fig.~\ref{fig:realistic_phaseshifts}\, shows the phase difference $2 \theta \equiv \arg{\psi_\text{out}}-\arg{\psi_\text{in}}$ between incoming and outgoing momentum components as a function of momentum and verifies our approximation by a linear function for sufficiently narrow momentum distributions.
Here, we have chosen the parameters $p_0/ \hbar= 10 \cdot p_\text{rec}/\hbar= 8.05 \cdot 10^{7}\, \mathrm{m}^{-1}$, $\lambda=10^{-8}\,\mathrm{m}$, $\kappa =\sqrt{20} \cdot p_0/\hbar $, and $\Delta p_0 = 0.05 p_\text{rec}$ so that the incoming momentum is ten times the single-photon recoil of \textsuperscript{87}Rb atoms $p_\text{rec}$ in a typical Raman setup, and the decay length of the exponential potential $\lambda \cong  10^{-8}\, \mathrm{m}$ is significantly smaller than the spatial extension of such an interferometer. The momentum width is of the order of typical experiments with ultracold quantum gases.
For the sake of simplicity, we use the coordinate system where $s= 0$.

In this regime, the mirror operator from~Eq.~\eqref{eq:mirror_operator} can be applied if we use the effective position of the mirror defined by parameters of the potential.
The effective position of the potential can be adjusted by changing the physical position $s$ such that the condition for a closed interferometer is fulfilled.
We note that there is an additional constant phase offset $2\theta_0$, which we can choose to be the same for both arms by adjusting the potential strength and position.

\section{Challenges}
\label{sec:Challenges}
When implementing a specular mirror for wave packets by an exponential potential, defects in the mirror may cause a loss of coherence~\cite{Esteve04}.
In addition, wave front distortions arise in the quasispecular regime~\cite{Henkel97} and are in an SMI intrinsically caused by gravity.
To minimize this effect, the amplitude of the exponential potential needs to be much larger than the gravitational potential difference across the wave packet.
At the same time, this condition limits the expansion time of the atomic cloud, and therefore, the time the atoms spend in the SMI.

To perform the experiment using initially located wave packets with narrow momentum distributions, it might be beneficial to resort to Bose-Einstein condensates, and to Bragg diffraction which has the additional advantage that the clock phase $-2\omega T$ in~Eq.~\eqref{eq:P_SMI} vanishes.
In addition, a Bragg configuration would allow for a superposition of two internal states on each branch as proposed in Ref.~\cite{Zych11}.

Furthermore, employing large-momentum-transfer techniques~\cite{Chiow11} rather than two-photon beam splitters would enable longer interferometer times and higher relative velocities with respect to the mirrors at reflection.
In fact, their use is crucial to prevent the upper wave packet from starting to fall down before it has been completely reflected, and to minimize wave packet distortions due to gravity during the reflection process.

\section{Conclusions}
\label{sec:Conclusions}
We have proposed a new geometry for atom interferometers with specular mirrors, where in contrast to an MZI the nonrelativistic signature of proper time is nonvanishing.
This effect is directly related to an asymmetry in the action of the specular mirrors on the atoms, and is independent of their specific implementation.
Indeed, a specular mirror and a diffractive momentum transfer give two different phase contributions.
In this sense, the action of the mirror plays an integral part of the interferometer.
Finally, we have introduced a mirror operator and we have shown that it can be associated with the asymptotic reflection of a particle from an evanescent field.

Because the effective visibility is the observable in Ref.~\cite{Zych11}, a good signal-to-noise ratio as well as a closed geometry together with a nonvanishing proper-time difference is of particular importance.
Since the SMI is both closed in phase space and has no intrinsic loss of particles like other configurations~\cite{Borde08}, it is an ideal test-bed to measure proper-time effects in atom interferometers.

\begin{acknowledgments}

We thank H. Abele, S. Abend, Ch. J. Bord\'{e}, {\v{C}} Bruckner, C. Feiler, D. Heim, M. Kasevich, S. Kleinert, H. Lemmel, S. Loriani, D. Petrascheck, I. Pikovski, H. Rauch, S. Reynaud, D. Schlippert, C. Ufrecht, S. Werner, P. Wolf, W. Zeller, M. Zimmermann, and M. Zych for many fruitful discussions. We are also grateful to H. M{\"u}ller for bringing Ref.~\cite{Borde93} to our attention. 
The presented work is supported by the CRC 1227 DQ-mat, CRC 1128 geo-Q, QUEST-LFS, and the German Space Agency (DLR) with funds provided by the Federal Ministry of Economic Affairs and Energy (BMWi) due to an enactment of the German Bundestag under Grant No. DLR 50WM1552-1557 and 50WM1641.
The work of IQ\textsuperscript{ST} is financially supported by the Ministry of Science, Research and Arts Baden-W\"urttemberg.
WPS thanks Texas A{\&}M University for a Faculty Fellowship at the Hagler Institute for Advanced Study at Texas A{\&}M University, and Texas A{\&}M AgriLife for the support of this work.
DMG is grateful to the Alexander von Humboldt-Stiftung for a Wiedereinladung which made the start of this work possible, and to the John Templeton Foundation for the grant \#21531, which allowed its completion.

\end{acknowledgments}

\appendix

\section{Interference pattern in the SMI}
\label{sec:Appendix}
According to Eq.~\eqref{eq:state_SMI}, the external degree of freedom of the atoms exiting an SMI in the ground state can be described by the state
\begin{equation}
    \ket{\psi_\text{SMI}^{(g)}} = \bra{g} \hat{R}^{(B)}_{-}(2T)\hat{U}(T)\hat{M}(\zeta_e,\zeta_g)\hat{U}(T)\hat{R}^{\left(B\right)}_{+}(0)\ket{\Psi_\text{in}},
\end{equation}
where the explicit form
\begin{equation}
    \hat{R}^{(B)}_{\pm}(t) \equiv \frac{1}{\sqrt{2}}\left[ \mathbbm{1}_\text{int}
      -\ii 
    \left( \ee^{ \pm \ii (k \hat{z} + \phi_t)}\ket{e}\!\bra{g} +
        \text{h.c.}
    \right) \right]
\end{equation}
of the Raman beam splitter can be found from Eq.~\eqref{eq:Raman}.

We assume that the atom is initially in the state $\ket{\Psi_\text{in}}\equiv \ket{e}\ket{\psi}$, where $\ket{\psi}$ is describes the external degree of freedom and $\ket{e}$ denotes the excited state of the atom.
When we calculate the action of the first beam splitter on the excited state, we find
\begin{equation}
    \hat{R}^{\left(B\right)}_{+}(0)\ket{e}= \frac{1}{\sqrt{2}}\left[ \ket{e} -\ii   \ee^{ - \ii (k \hat{z} + \phi_0)}\ket{g}\right].
\end{equation}
Because we postselect on the population in the ground state, we also make use of the product
\begin{equation}
  \bra{g}  \hat{R}^{\left(B\right)}_{-}(2T)= \frac{1}{\sqrt{2}}\left[ \bra{g} -\ii   \ee^{  \ii (k \hat{z} + \phi_{2T})}\bra{e}\right].
\end{equation}
With these two relations it is easy to see that the final state 
\begin{align}
\begin{split}
        \ket{\psi_\text{SMI}^{(g)}} = \frac{\ii}{2}&\left[\bra{g}\hat{U}(T)\hat{\Pi}(\zeta_g)\hat{U}(T)\ket{g}\ee^{-\ii (k \hat{z}+\phi_0)} \right.\\
        &\left.+\ee^{\ii (k \hat{z}+\phi_{2T})}\bra{e}\hat{U}(T)\hat{\Pi}(\zeta_e)\hat{U}(T)\ket{e}\right]\ket{\psi}
\end{split}
\end{align}
of the atom exiting the interferometer in the ground state consists of a superposition of the atom traveling along the lower path in the ground state \emph{and} the upper path in the excited state.

To describe the time evolution of the atom in the gravitational field, we make use of the fact that the internal degree of freedom is separable from the external one.
We therefore introduce the operator
\begin{equation}
\label{eq:internal_time_evolution}
    \hat{\mathcal{U}}\equiv \exp\left\lbrace-\ii \left[\hat{p}^2/(2m) + m g \hat{z} \right] T /\hbar \right\rbrace
\end{equation}
that acts on the latter, and use the phase factor $\exp[\mp \ii \omega T/2]$ when the atom propagates in the excited or ground state, respectively.

With this notation, the final state reads
\begin{align}
\begin{split}
        \ket{\psi_\text{SMI}^{(g)}} = \frac{\ii}{2}&\left[\ee^{-\ii(\phi_0-\omega T)}\hat{\mathcal{U}}\hat{\Pi}(\zeta_g)\hat{\mathcal{U}}\ee^{-\ii k \hat{z}} \right.\\
        &\left.+\ee^{\ii (\phi_{2T}-\omega T)}\ee^{\ii k \hat{z}}\hat{\mathcal{U}}\hat{\Pi}(\zeta_e)\hat{\mathcal{U}}\right]\ket{\psi}.
\end{split}
\end{align}
From this expression we find the interference pattern described by the probability $P_\text{SMI}^{(g)}\equiv \langle\psi_\text{SMI}^{(g)}|\psi_\text{SMI}^{(g)}\rangle$ and arrive at
\begin{align}\label{eq.probability_app}
     P_\text{SMI}^{(g)} =& \frac{1}{2}+\frac{1}{4} \ee^{\ii (\phi_+ -2\omega T)}  \bra{\psi}   \hat{\mathcal{O}} \ket{\psi} +\text{ c.c. },
\end{align}
where we have introduced the operator
\begin{equation}
     \hat{\mathcal{O}} \equiv \ee^{\ii k \hat{z}} \hat{\mathcal{U}}^\dagger \hat{\Pi}^\dagger(\zeta_g) \hat{\mathcal{U}}^\dagger \ee^{\ii k \hat{z}} \hat{\mathcal{U}} \hat{\Pi}(\zeta_e) \hat{\mathcal{U}}.
\end{equation}
To simplify $\hat{\mathcal{O}}$, we decompose the time-evolution operator from Eq.~\eqref{eq:internal_time_evolution} with the help of the Baker-Campbell-Hausdorff and Zassenhaus identities into the two equivalent formulae
\begin{subequations}
\begin{equation}
    \hat{\mathcal{U}}= \ee^{-\ii mg\hat{z}T/\hbar } \ee^{- \ii  \hat{p}^2 \label{eq:U_1} T/(2m\hbar)}\ee^{\ii g T^2 \hat{p}/(2\hbar)} \ee^{-\ii m g^2 T^3 /(6 \hbar) }
\end{equation}
and
\begin{equation}
    \hat{\mathcal{U}}= \ee^{- \ii  \hat{p}^2 T/(2m\hbar)}\ee^{-\ii g T^2 \hat{p}/(2\hbar)} \ee^{-\ii mg\hat{z}T/\hbar }  \ee^{-\ii m g^2 T^3 /(6 \hbar) } \label{eq:U_2}.
\end{equation}
\end{subequations}

We calculate the operator product $\hat{\mathcal{U}} \hat{\Pi}(\zeta) \hat{\mathcal{U}}$ and sandwich the parity operator between the representations of the time evolution given in Eq.~\eqref{eq:U_1} and Eq.~\eqref{eq:U_2}.
The action of the momentum operators on the momentum representation of the parity operator, see Eq.~\eqref{eq:parity}, can be performed trivially and the operators $\exp[-\ii mgT \hat{z}/\hbar]$ lead to an additional shift in momentum.
In total, the product takes the form
\begin{equation}
\label{eq:parity_time}
\begin{split}
    \hat{\mathcal{U}} \hat{\Pi}(\zeta) \hat{\mathcal{U}}= \int \!\! \dd p \,& \exp\left[-\frac{\ii}{\hbar}  \left(\frac{p^2 T}{m} +p g T^2 - 2 p \zeta\right)\right] \\
    & \times \ket{- p-mgT}\!\bra{p+mgT} \ee^{-\ii \frac{ m g^2 T^3 }{3\hbar}} 
\end{split}
\end{equation}
in momentum representation.

We evaluate the second part of the operator $\hat{\mathcal{O}}$ given by the product
$
\ee^{\ii k \hat{z}} \hat{\mathcal{U}}^\dagger \hat{\Pi}^\dagger(\zeta_g) \hat{\mathcal{U}}^\dagger \ee^{\ii k \hat{z}}
$ by taking the Hermitian conjugate of Eq.~\eqref{eq:parity_time} and shifting the momentum with the operators $\exp[\ii k \hat{z}]$.
Hence, the projection operator in the integral reads
$\ket{ p+mgT+\hbar k}\!\bra{-p-mgT-\hbar k} $.

Multiplying these two operator sequences and using the scalar product of the two projection operators, the operator $\hat{\mathcal{O}}$ reduces to
\begin{align}
\begin{split}
    \hat{\mathcal{O}} = \int \!\! \dd p\, \ee^{  \ii  \left(\frac{\hbar k^2 T}{m} +2 k \zeta_g -k g T^2 + \frac{p Z}{\hbar}\right) } \ket{ p+mgT}\!\bra{p+mgT},
\end{split}
\end{align}
where we have defined the classical separation $Z \equiv \zeta_e-\zeta_g - k T/m$ of the two mirrors.

Shifting the variable of integration by $mgT$, introducing the momentum operator $\hat{p}$ and making use of the completeness relation $\int\!\! \dd p \ket{ p}\!\bra{p}= \mathbbm{1}_\text{ext} $ of the momentum eigenstates, the operator
\begin{align} \label{eq.overlap_appendix}
    \hat{\mathcal{O}} = \ee^{  \ii  \left(\hbar k^2 T/m +2 k \zeta_g -k g T^2 \right) } \ee^{\ii 2 (\hat{p}-mgT)Z/\hbar}
\end{align}
consists of phase factors and an additional displacement operator in position.
When we substitute Eq.~\eqref{eq.overlap_appendix} into Eq.~\eqref{eq.probability_app}, we find exactly the form of the interference pattern used in Eq.~\eqref{eq:P_SMI} in the main part of the article.

\bibliography{50_bib}
\end{document}